\documentclass[12pt,preprint]{aastex}

\shorttitle{Mid-IR view of M31}
\shortauthors{Barmby et al.}

\newcommand{\mic}{~$\mu$m\ }
\newcommand{\mice}{~$\mu$m}
\newcommand{\mjsr}{~MJy~sr$^{-1}$}

\begin{document}

\title{Dusty waves on a starry sea: the mid-infrared view of M31}

\author{
P. Barmby\altaffilmark{1}, 
M.L.N. Ashby\altaffilmark{1},
L. Bianchi\altaffilmark{2},
C.W. Engelbracht\altaffilmark{3},
R.D. Gehrz\altaffilmark{4},
K.D. Gordon\altaffilmark{3}, 
J.L. Hinz\altaffilmark{3},
J.P. Huchra\altaffilmark{1},
R.M. Humphreys\altaffilmark{4},
M.A. Pahre\altaffilmark{1},
P.G. P\'erez-Gonz\'alez\altaffilmark{3}, 
E.F. Polomski\altaffilmark{4},
G.H. Rieke\altaffilmark{3},
D.A. Thilker\altaffilmark{4},
S.P. Willner\altaffilmark{1},
C.E. Woodward\altaffilmark{4}
}

\altaffiltext{1}{%
Harvard-Smithsonian Center for Astrophysics, 60 Garden St., Mailstop 65, Cambridge, MA 02138;
pbarmby, willner, mashby, huchra, mpahre@cfa.harvard.edu
}
\altaffiltext{2}{%
Astrophysical Sciences, The Johns Hopkins University, 3400 North Charles St., Baltimore, MD 21218;
bianchi, dthilker@pha.jhu.edu
}
\altaffiltext{3}{%
Steward Observatory, The University of Arizona, 933 N. Cherry St., Tucson, AZ 85721;
cengelbracht, kgordon, jhinz, pgperez, grieke@as.arizona.edu
}
\altaffiltext{4}{%
Astronomy Department, School of Physics and Astronomy, University of Minnesota, 
 116 Church St, SE, Minneapolis, MN 55455;
gehrz@astro.umn.edu, roberta@aps.umn.edu, elwood@astro.umn.edu, chelsea@astro.umn.edu
}

\begin{abstract}
Mid-infrared observations of the Andromeda galaxy, M31, obtained
with the Infrared Array Camera on board the Spitzer Space Telescope
are presented.
The image mosaics cover areas of approximately $3\fdg7 \times 1\fdg6$
and include the satellite galaxies M32 and NGC~205.
The appearance of M31 varies dramatically in the different
mid-infrared bands, from the smooth bulge and disk of the old stellar
population seen at 3.6\mic to the well-known `10 kpc ring' dominating
the 8\mic image. The similarity of the 3.6\mic and optical isophotes
and nearly constant optical-mid-infrared color over the inner 
400\arcsec\ confirms that there is no significant extinction at optical
wavelengths in M31's bulge. The nuclear colors indicate the presence
of dust but not an infrared-bright active galactic nucleus.
The integrated 8\mic non-stellar luminosity implies a star
formation rate of $0.4 M_{\sun}$~yr$^{-1}$, consistent with other indicators
that show M31 to be a quiescent galaxy.
\end{abstract}


\section{Introduction}

The proximity of the Andromeda galaxy, M31, has long made it a prime target for studies of 
galaxy structure and stellar populations. One technique has been to isolate particular
populations: for example, studies of luminous stars \citep{hmf90}
showed that M31 has a lower massive star formation rate than M33 and the LMC.
Global studies of particular components
have also been valuable: H~I maps provide estimates of the mass distribution
\citep{bt04} and have recently revealed a population of high-velocity clouds \citep{thil04}.
{\it ISOPHOT} \citep{haas98}, {\it MSX} \citep{kraemer02}, and {\it IRAS} \citep{habing84} observations showed a 
bright ring of infrared emission with radius 10~kpc coincident with many of the H~II regions.

Mapping the entire disk of M31 at mid-infrared wavelengths allows {\em both} local
and global studies of the galaxy.  Observations with the Infrared Array Camera
\citep[IRAC;][]{irac} on the {\it Spitzer Space Telescope}
simultaneously trace the dust in the 
spiral arms at 8\mic and the oldest stars in the disk and bulge at 3.6 and 4.5\mic
without the complicating extinction or distance effects that make such studies in the Milky Way difficult.  
IRAC observations of M31 are complemented by deep data
now available at many other wavelengths. They also complement {\it Spitzer} studies
of other nearby galaxies.

This paper presents an initial look at the IRAC observations of M31, focusing on the
surface brightness profiles and extended emission. Companion papers discuss 
longer-wavelength MIPS observations of M31 \citep{gordon06}, IRAC and MIPS observations of the M31 satellite 
galaxy NGC~205 \citep{marleau06}, and the implications of the galaxy's morphology
as seen in 8\mic non-stellar emission \citep{block06}.
A distance to M31 of 783~kpc \citep{sg98}
is assumed throughout. All magnitudes are on the Vega system, using the calibration
given by \citet{reach05}.

\section{Observations and Data Reduction}

The IRAC observations of M31 were taken as part of {\it Spitzer} General Observer 
program 3126 in 2005 January and August.%
\footnote{A large solar proton event made most of the January data unusable;
re-observations were made in August.}
Fifteen Astronomical Observation Requests (AORs) were used to map a region approximately 
$3\fdg7 \times 1\fdg6$ (chosen to match the {\it Spitzer}/MIPS observations
made as part of program ID 99), with an extension to the NW to include NGC~205.
The central $1\fdg6 \times 0\fdg4$ was covered by three AORs, each having two 
12-second frames per position. The outer regions were covered by two AORs each with two dithered 
30-second frames per position. This mapping strategy ensured that
observations of each position in the galaxy were separated by at least 2.5 hours,
allowing efficient asteroid rejection in data processing.
The complete dataset consists of 3000 individual images in each of the IRAC channels.

Data reduction began with the Basic Calibrated Data (BCD) produced by versions 11 (for the January data) 
or 12 (the August data) of the {\it Spitzer} Science Center (SSC) Pipeline. 
A `delta dark' offset correction was applied to the 12-second 5.8\mic frames to correct for
the first-frame effect, followed by
use of the `artifact corrector' software developed by S. Carey,
which attempts to remove the electronic effects caused by bright stars.
The remaining processing steps were carried out using the 
``IRAC\_proc''\footnote{IRAC\_proc was developed by M. Schuster, M. Marengo and B. Patten at 
the Smithsonian Astrophysical Observatory.} software with the 2005 September 5 version of the SSC 
MOPEX package.
The `overlap correction' necessary to make the sky background consistent between frames 
was computed and applied. Median-combined mosaics of the full dataset were created, then projected
into the reference frame of each input BCD image to create `source maps'. These source maps were used
as the input `uncertainty images' for creating the final photometric mosaics --- this step prevents the
outlier rejection used by MOPEX from rejecting pixels in bright sources and thereby biasing the photometry.
The final mosaics were created with a pixel scale of 0\farcs86, sub-sampling the native IRAC
pixel scale by a factor of $\sqrt{2}$ to better sample the point spread function 
(FWHM $\sim1\farcs9$).
The sky background was removed using the {\sc iraf} task {\sc imsurfit} to
model the background as a polynomial surface sampled by `blank sky'
regions near the edge of the mosaics. After subtraction, the mean image counts in small areas
within these regions were computed, and the standard deviation of these means
used to estimate the photometric uncertainties due to the background.

The final step in mosaic production was photometric calibration.
Although calibration of point source photometry varies slightly over the IRAC arrays' field of view 
\citep{reach05}, we did not correct for this since
doing so would have corrupted the extended emission.
The calibration of IRAC is known to differ between point and extended sources, particularly at 5.8
and 8\mic where a significant fraction of point-source light is scattered on scales comparable
to the array size. We used the results of recent work by T. Jarrett (priv. comm.)
to correct from point to extended source calibration in the M31 mosaics by multiplying 
the images by (0.91, 0.94, 0.66, 0.74) in the four IRAC bands, respectively
[these values differ slightly from those given in \citet{reach05}].
The standard deviations of the background levels are about 0.04\mjsr\ in all bands except 5.8\mice,
where the background subtraction was more difficult, and the noise was
0.1\mjsr (1$\sigma$).
These correspond to limiting surface brightness levels of $21.2, 20.9, 19.0$, and 
19.6~mag~arcsec$^{-2}$ in the four IRAC bands.

\begin{figure}
\epsscale{0.80}
\plotone{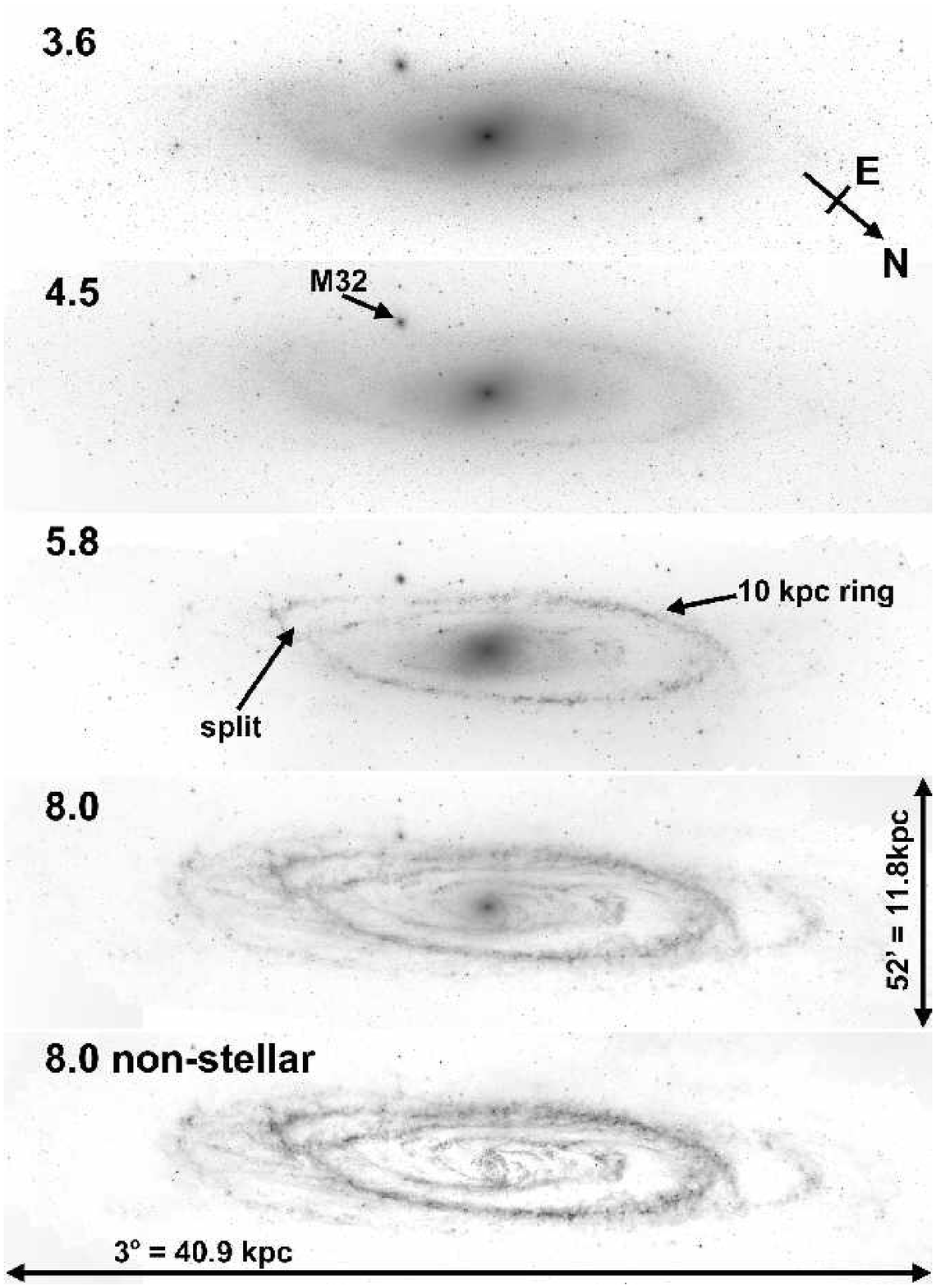}
\caption{(Plate 1) Mid-infrared mosaics of M31 as seen in the IRAC bands. 
The grayscales are approximately logarithmic with intensity. From
top to bottom: 3.6\mice, 4.5\mice, 5.8\mice, 8\mice, and 8\mic `non-stellar'
(produced by subtracting the expected stellar contribution to this band,
a scaled version of the 3.6\mic image).
All images are centered on the nucleus and cover an area of $3\fdg0 \times 0\fdg87$.
NGC~205 was observed with IRAC but is not shown here \citep[see][]{marleau06}.
\label{m31_bw}}
\end{figure}

The final of mosaics are shown in Figure~\ref{m31_bw}.
As with other recent works on IRAC observations of nearby galaxies 
we assumed that the 3.6\mic emission is purely due to stars, and used this to construct
maps of the `non-stellar' emission at 5.8 and 8\mice. The stellar emission at the two wavelengths
is assumed to correspond to 0.399 and 0.232 $\times$ the 3.6\mic emission \citep{helou04}.
The 8\mic non-stellar image is shown in Figure~\ref{m31_bw}.

\section{Morphology and surface brightness profiles}

M31's appearance changes dramatically over the IRAC bands, in a similar way to 
other spiral galaxies such as M81 \citep{irac_m81} and NGC~300 \citep{helou04}.
The 3.6\mic view is similar to the $K_s$-band image presented
in \citet{2mass_lga}: dominated by the bulge, with the spiral arms and disk relatively
faint.%
\footnote{The 2MASS Large Galaxy Atlas image of M31 is suspected of having an overly-faint 
disk due to problems with background subtraction (T. Jarrett, pers. communication). 
We do not compare the IRAC quantitative measurements to those given by \citet{2mass_lga},
instead estimating M31's $K-[3.6]$ color by M81's value of 0.3 \citep{irac_m81}.} 
At 8\mice, M31's
bulge is much less prominent, and the spiral arms and the 10~kpc ring are the dominant features.
The outer (14~kpc) ring detected at far-infrared wavelengths \citep{haas98,gordon06} 
and tentatively detected with {\it MSX} \citep{kraemer02} is also seen in the 8\mic image.
The 8\mic image --- particularly the non-stellar version --- looks extremely similar to 
the 24\mic image presented by \citet{gordon06}, showing the same split on the
southwest side of the 10~kpc ring that \citet{gordon06} attribute to interactions with M32. 
The `hot spots' seen near the nucleus at 160\mic are also prominent in the 8\mic
non-stellar image, where they coincide with the ends of spiral arms.

Disk images $3\fdg3\times0\fdg9$ were extracted from the mosaics
and used to derive IRAC surface brightness profiles of M31.
Peaks of bright stars, a 1\arcmin-radius region around M32, and image
regions with no coverage were all masked from the profile measurements.
Profiles were measured using two methods: making 20\arcsec-wide cuts along the disk
major and minor axes using the {\sc iraf} task {\sc pvector}, and fitting
elliptical isophotes to the images using the {\sc ellipse} task.
Because {\sc ellipse} can be unstable at
low surface brightnesses, the isophote shape parameters (ellipticity, 
position angle, center) were constrained to be constant beyond the edge of the visible disk
at semi-major axes $>0\fdg88$. The isophote fit at this distance
has an ellipticity ($\epsilon=1-b/a$) of 0.73 and position angle of
38\arcdeg. Profiles in the three longer-wavelength bands were measured on the
ellipses fit to the 3.6\mic image, so that colors could be measured at the
same spatial locations.

\begin{figure}
\plotone{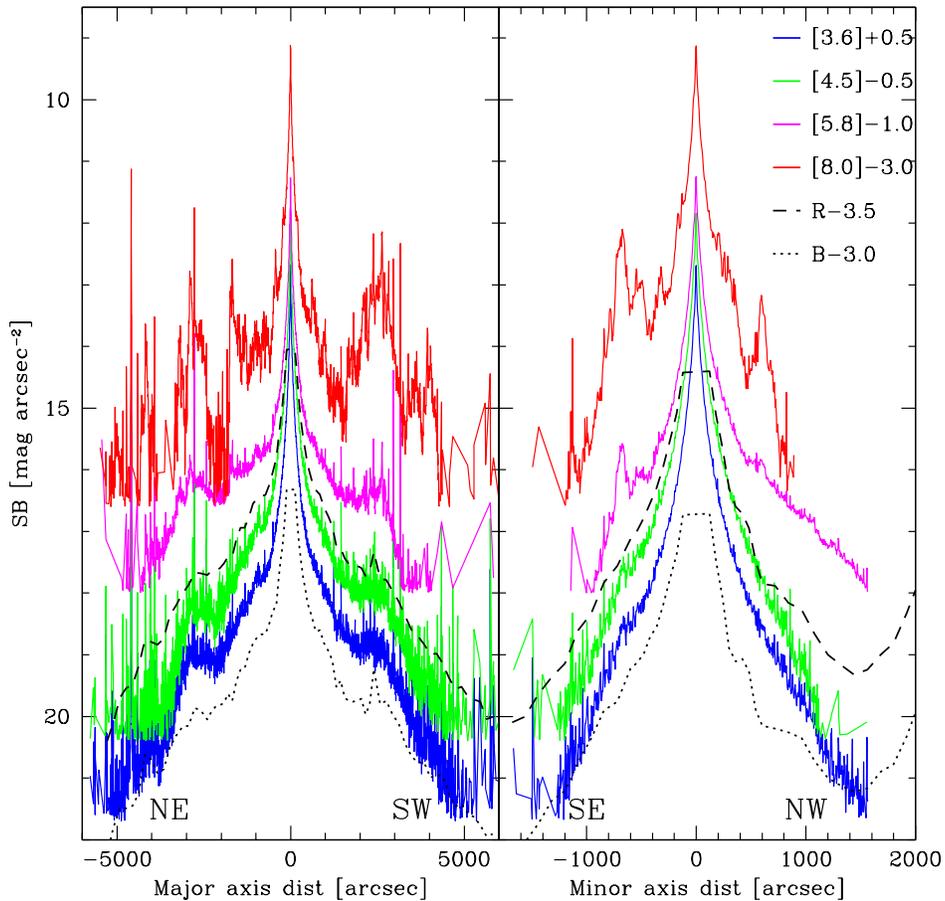}
\caption{Surface brightness profiles measured along the M31 disk major (left)
 and minor (right) axes.
 Profiles have been offset for clarity as indicated in the legend.
 The mid-infrared profiles are compared to optical profiles from \citet{wk87},
 which are plotted as flat in the center where no data were available. The upturn in the
 northwest minor axis profiles of \citet{wk87} is due to NGC~205, not included
 in the IRAC images used here.
\label{fig_axes}}
\end{figure}

The relative prominence of the M31 disk and bulge in the four IRAC bands
is reflected in the major and minor axis profiles, shown in Figure~\ref{fig_axes}.
Out to a semi-major axis of $\sim400$\arcsec\ (approximately
the edge of M31's bulge), all four bands have very similar profiles.
Beyond 400\arcsec, the 5.8\mic and 8.0\mice-band profiles decline
more slowly and show more variability due to the clumpy nature
of the dust emission. The 10~kpc ring is visible in all of the major-axis
profiles; the 8.0\mic profile in particular appears highly asymmetric 
because of the split in the ring. The minor axis profiles are 
smoother because they sample mainly the bulge. Both major and minor axis
profiles are quite similar in shape to the optical profiles presented by \citet{wk87},
indicating that dust extinction does not have a major effect on the optical profiles.
The nucleus of M31 is smaller than the IRAC PSF, and does not 
separate from the bulge emission in the radial profile. 

\begin{figure}
\plotone{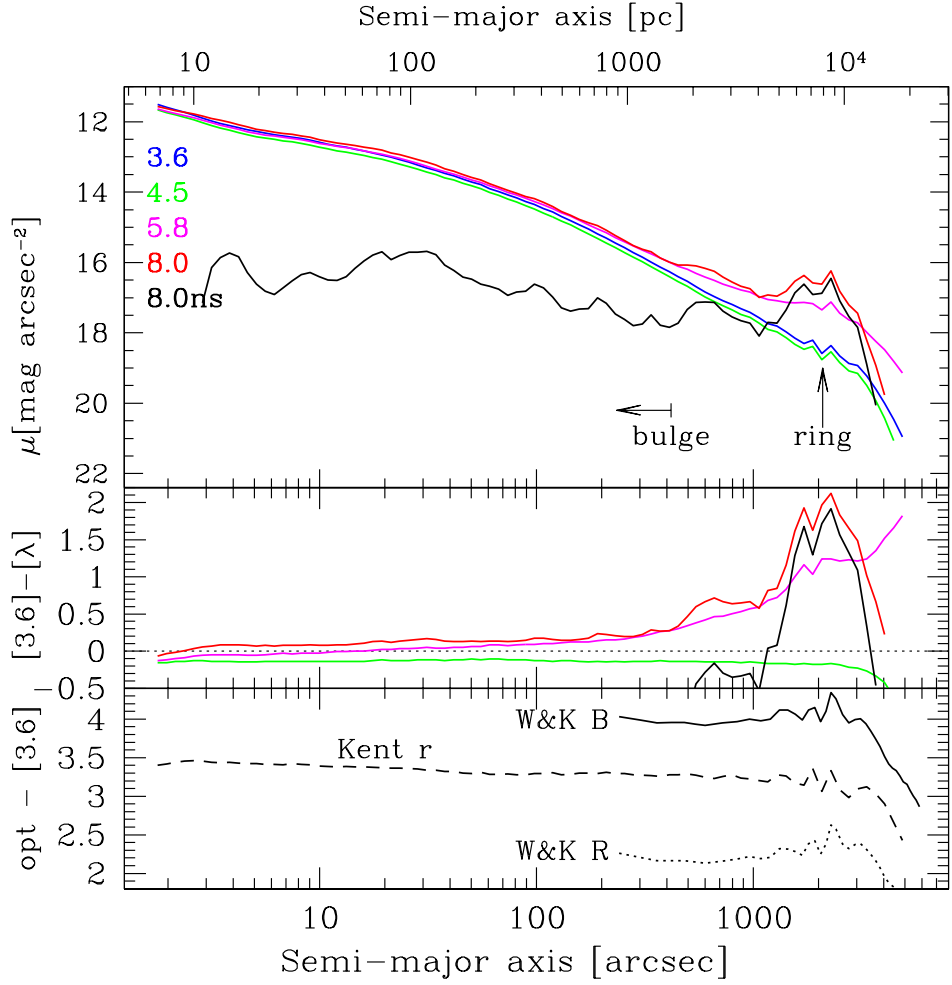}
\caption{
Mid-infrared surface brightness and color profiles of M31
derived from {\sc iraf/ellipse} isophotes fitting. All data use the extended
source calibration, as described in the text.
`8.0ns' refers to the `non-stellar' 8\mic image.
(Top): IRAC surface brightness profiles.
(Middle): IRAC color profiles from the same data.
(Bottom): Optical-mid-infrared color profiles, using optical data from
\citet{kent87} and \citet{wk87}. The large offset between colors using $r$ and $R$
is likely due to problems with the zero-point calibration of the \citet{wk87} data.
\label{fig_colors}}
\end{figure}

The color profiles of M31 shown in Figure~\ref{fig_colors} also depict the
disk/bulge separation. The median isophote colors within 400\arcsec\ are
$[3.6]-[4.5] = -0.14$, $[3.6]-[5.8]=0.03$ and $[3.6]-[8.0]=0.13$.
The $[3.6]-[4.5]$ color is consistent with theoretical expectations for M0 
giant stars \citep[see][]{pahre04}; the $[3.6]-[5.8]$ and $[3.6]-[8.0]$
colors are slightly redder, presumably because of dust emission.
At larger distances, the disk dominates the light and the colors become redder,
particularly at the location of the 10 kpc ring.
Comparing the 3.6\mic profile to the $r$-band optical 
profile of \citet{kent87} shows that the $r-[3.6]$ color is nearly constant 
over the central 400\arcsec, with a luminosity-weighted mean of $r-[3.6]=3.41$~mag.
This corresponds to $r-K\approx3.1$, reasonably consistent with the predictions of
\citet{mar05}.
The colors of the M31 nucleus (measured in a 2\farcs4-radius aperture,
with the IRAC point source calibrations and aperture corrections applied)
are $[3.6]-[4.5] = -0.15$, $[3.6]-[5.8]=0.19$ and $[3.6]-[8.0]=0.51$.
The red colors at the longer wavelengths suggest that the nucleus
has more dust than the bulge, but are not the very red colors characteristic
of an AGN \citep[e.g.][]{stern05}.

Table~\ref{tab_mag} gives integrated flux densities and magnitudes of M31 as measured in the largest 
elliptical isophote ($a=$1\fdg8). The uncertainties in these values include the
effects of background variation and a
10\% uncertainty in the absolute extended source calibration.
The IRAC flux densities are consistent with the COBE/DIRBE measurements
at 3.5 and 4.9\mic \citep[245 and 128 Jy with 20\% uncertainties;][]{ons98} 
and the {\it MSX} A-band (8.3\mice) value \citep[$159\pm32$ Jy;][]{gordon06}.
Near-to-mid-infrared colors are another way to check our photometry:
\citet{amh80} measured $H=1.42$ in a 1554\arcsec\ diameter circular aperture centered on M31.
In the same aperture the 3.6\mic magnitude is 0.84, giving $H-[3.6]=0.58$.
Using M81's $K_s-[3.6]\approx0.3$ \citep{irac_m81} and the mean $H-K_s= 0.27$
measured for early-type (S0/Sa/Sb) spirals by \citet{2mass_lga}, one
would predict $H-[3.6]=0.57$, in very good agreement with our measurement.

A model light distribution consisting of an $r^{1/4}$ bulge and exponential disk 
were fit to the 3.6\mic semi-major axis surface brightness (SB) profile.
The structural parameters fitted to the disk are sensitive to inclusion of the ring 
at semi-major axis lengths $2000 \leq a \leq 3500$~arcsec,
while the bulge effective radius is
sensitive to the inner fitting cutoff due to deviations from the idealized $r^{1/4}$ model
at $a < 100$~arcsec  \citep[see][]{kent83}.
The sky value for the image was determined at isophotes roughly corresponding to $3.5$ disk scale lengths.
As there is still some galaxy emission at this radius, our sky value is likely
over-estimated and is probably the dominant cause of the
significant reduction in the SB at $a > 4000$~arcsec (smaller, but still
significant, errors will be present in the profile at $a < 4000$~arcsec).
We chose the fitting regions of $100 \leq a \leq 2000$ and $3500 \leq a \leq 4000$~arcsec
to minimize these effects, and for consistency with previous work.

The resulting model has (bulge,disk) half-light semi-major axis lengths of $(10.3\pm 0.4,44.7\pm 0.6)$~arcmin,
average ellipticities near to those scale lengths of $(0.48\pm 0.02,0.68\pm 0.04)$,
total magnitudes $(0.84\pm 0.13,0.58\pm 0.06)$, and RMS of $0.02$~mag.
The corresponding circularized effective radii are $(7.5\pm 0.3,25.2\pm 0.7)$~arcmin.
This disk major axis scale length of $6.08\pm 0.09$~kpc compares well with the longest wavelength
($R$-band) value of 5.9~kpc measured by \citet{wk88}.
The ratio of the bulge effective radius of $1.70\pm 0.07$~kpc to disk scale length 
place M31 on the upper envelope of local spiral galaxies \citep{dej96}.
The fitted bulge and disk models have a flux ratio $B/D = 0.78 \pm 0.11$.
This value is biased high relative to the true ratio, however, because the
bulge model is too bright at $a < 100$~arcsec and the disk model is too faint at the location
of the 10~kpc ring.
\citet{wk88} calculate a similar value in the $R$-band.

The model-fitting allows an estimate of M31's total magnitude at 3.6\mic through
extrapolation of the disk model to infinite radius.
The magnitude within $a < 4000$~mag is 0.19~mag, 
the extrapolated disk from 2.5 scale lengths to infinity adds an additional 0.38~mag,
and the sky subtraction errors on the SB profile within this isophote
result in an additional 0.15~mag.
The implied total magnitude of M31 is therefore $-0.34$~mag. We have chosen not to 
quote this model-dependent value in Table~1 for consistency with previous work and our
measurements at other wavelengths, but it is consistent with the measurement in the 1\fdg8 isophote
given there when similar corrections are applied.
The errors on the total magnitude of M31 are dominated by the systematic error
of sky subtraction on this large galaxy, and less so on the extrapolation of the disk emission.

The 3.6\mic luminosity of M31 can be used to estimate the galaxy's stellar mass. 
Using a typical value of $B-R=1.5$ from \citet{wk87}
and the color-dependent mass-to-light ratios
of \citet{bdj01}, an estimated $M/L_K$ for M31 is about 0.8 in solar units.
Using $K-[3.6]=0.3$  and multiplying by $M/L_K$, the resulting stellar mass is
$1.1\times10^{11} M_{\sun}$. Recent dynamical mass estimates of the M31 disk+bulge 
\citep[e.g., ][]{wps03,gee06} 
are $\sim10^{11}M_{\sun}$, in reasonable agreement with this
derived stellar mass. 
Asymptotic giant branch (AGB) stars can be important contributors to 
the total near-infrared luminosities of galaxies \citep{mar05}.
However, the most luminous AGB stars are predicted by \citet{gro06} to have
much redder colors ($0.5<[3.6]-[4.5]<1$) than observed for M31, so
we conclude that they do not dominate M31's 3.6\mic luminosity.

The 8\mic non-stellar emission is dominated by the 7.7\mic polycyclic aromatic hydrocarbon (PAH)
band, which is a useful though imperfect star formation rate (SFR) indicator. It is therefore
instructive to compare the M31 8\mic luminosity with other star formation
indicators, such as H$\alpha$, radio, and total infrared luminosities and with
these values for other galaxies. 
\citet{wu05} derived a calibration for SFR
as a function of 8\mic non-stellar luminosity $\nu L_{\nu}[8\mu{\rm m(dust)}]$ 
(hereafter $L_8$)
using correlations between $L_8$ and $L({\rm H}\alpha)$ and $L(1.4 {\rm GHz})$.
The IRAC 8\mic non-stellar flux density measured 
for M31 corresponds to a luminosity of $\log(L_8/L_{\sun})={8.8}$.
The \citet{wu05} calibration yields an SFR of $0.4 M_{\sun}$~yr$^{-1}$.

It is possible to compare M31's 8\mice-derived SFR with SFRs from
other indicators, but since all such indicators have calibration uncertainties
[e.g., the \citet{wu05} 8\mice/SFR calibration is based on a 
small number of galaxies with a limited range of properties], we instead
compare the observed properties directly.
To estimate the $H\alpha$ luminosity of M31, we convert
the value given by \citet{dev94} to the 780~kpc distance, multiply by 0.8 to
account for [NII] contamination, and multiply by 3.4 \citep[as done by][]{wb94}
to correct for extinction. The resulting $L({\rm H}\alpha)=2.6\times 10^7 L_{\sun}$,
which predicts $\log(L_8/L_{\sun})=9.1$.
The 1.4~GHz radio flux density measured by \citet{bbh98}, $3.34\times 10^{20}$~W~Hz$^{-1}$,
yields a predicted $\log(L_8/L_{\sun})=8.4$.
Figure~12 of \citet{dale05} gives the ratio of $\nu f_{\nu}[8\mu{\rm m(dust)}]/f(3-1100\mu{\rm m})$
as a function of $f_{\nu}(70 \mu{\rm m})/f_{\nu}(160 \mu{\rm m})$ for SINGS galaxies. 
From \citet{gordon06}, M31 has $f(70)/f(160)=0.12$, with the corresponding 
$f(3-1100\mu {\rm m})=2.57\times10^{-10}$~W~m$^{-2}$, yielding a predicted $\log(L_8/L_{\sun})=9.0$.
The H$\alpha$ and far-infrared predictions are
in reasonably good agreement, and are consistent with the observed 
8\mic luminosity given the scatter in the relationships. The 8\mic luminosity
predicted from the radio flux is lower than observed, perhaps indicating that
some extended emission was resolved out of the radio maps.
All of the `star-formation' luminosities are at the low end of the galaxy distribution.
Despite its prominent star forming ring, M31 is a predominantly quiescent galaxy.

\acknowledgments

We thank the referee for helpful suggestions.
This work is based on observations made with the Spitzer Space Telescope, which is operated 
by the  Jet Propulsion Laboratory, California Institute of Technology under a contract with NASA. 
Support for this work was provided by NASA through an award issued by JPL/Caltech.

Facilities: \facility{Spitzer (IRAC)}

\clearpage

\clearpage

\clearpage

\clearpage

\begin{deluxetable}{crr}
\tablecaption{M31 mid-infrared flux densities and magnitudes\label{tab_mag}}
\tablewidth{0pt}
\tablehead{
\colhead{$\lambda$} & \colhead{$f_{\nu}$} &\colhead{$m_{\rm Vega}$} \\
\colhead{[$\mu$m]} & \colhead{[Jy]} & \colhead{}
}
\startdata
3.6 & $259\pm32$ &$  0.09\pm0.13$ \\
4.5 & $144\pm20$ &$  0.24\pm0.15$ \\
5.8 & $190\pm35$ &$ -0.55\pm0.20$ \\
8.0 & $151\pm21$ &$ -0.93\pm0.15$ \\
8.0 (non-stellar)  & $91\pm21$& \nodata
\enddata
\tablecomments{Measurements made in an elliptical aperture of semi-major
axis $a=$1\fdg8.}
\end{deluxetable}

\end{document}